\documentclass[aps,prb,twocolumn,reprint,superscriptaddress,amssymb,amsmath]{revtex4-2}
\usepackage{graphicx}
\usepackage{dcolumn}
\usepackage{bm, color}
\usepackage{amsfonts}
\usepackage[none]{hyphenat} 
\usepackage{bbm}
\usepackage{braket}
\usepackage{tabularx}
\usepackage{ulem}
\usepackage{amsmath}
\usepackage{newtxtext,newtxmath} 

\begin{document}
	
\title{Studies on Topological High-fold Degenerate Semimetal with Chiral Structure}
\author{Yan Wang}
\affiliation{Qingdao Institute for Theoretical and Computational Sciences, Shandong University, Qingdao 266237, China}
\author{Xiaosong Bai}
\affiliation{Qingdao Institute for Theoretical and Computational Sciences, Shandong University, Qingdao 266237, China}
\author{Wujun Shi}
\affiliation{School of Physical Science and Technology, ShanghaiTech University, Shanghai 200031, China}
\author{Wenjian Liu}
\affiliation{Qingdao Institute for Theoretical and Computational Sciences, Shandong University, Qingdao 266237, China}
\author{Qiunan Xu}
\email{xuqiunan91@sdu.edu.cn}
\affiliation{Qingdao Institute for Theoretical and Computational Sciences, Shandong University, Qingdao 266237, China}

\begin{abstract}
In recent years, a type of topological semimetals (TSMs) that can host new fermions with high-fold degeneracy has attracted considerable interest. Among them, ones with chiral structrue particularly catch our attention. Such chiral high-fold degenerate semimetals always have a larger topological charge and longer Fermi arcs which bringing about some special properties. In this work, we found 147 chiral materials with exotic fermions near Fermi level by high-throughput calculation and screening.  We selected some typical examples to analyse its topological properties such as topological surface states (TSSs) and Berry curvature. Our results are helpful to provide a promising platform for exploring the physical properties of chiral fermions and application of chiral TSMs. 

\end{abstract}

\maketitle

	
\section{Introduction}
\setlength{\parindent}{2em}
\vspace{-0.2cm}
With the deepening of theoretical and experimental research on topological insulators (TIs), the discussion on TSMs, which can be regarded as a nontrivial extension of TIs in metal materials, 
has emerged as a prominent topic in condensed matter physics.~\cite{TI_TSM_Hasan_2015, WSM_FermiArcs_Hasan_2016, TSM_3Dsolids_2018, WSM_TaAs_2015, WSM_TypeII_2015}. Among these, high-fold degenerate semimetals represent a new class of TSMs discovered in recent years. Compared with traditional Dirac semimetals and Weyl semimetals, TSMs can host exotic massless fermionic excitations with higher-fold degeneracy. The existence of these new fermions benefits from the absence of constraints imposed by Poincaré symmetry. Theoretical predictions and experimental observations have identified three-~\cite{HSM_3fold_ZrTe_2016, TSM_3fold_2016, TSM_3fold_2019, HSM_3fold_TaN_2016}, four-~\cite{HSM_4fold_2012}, six-~\cite{HSM_CoSi_2017, TSM_PtGa_2020, HSM_RhSi_2017}, and eight-fold~\cite{HSM_8fold_DoubleDirac_2016, HSM_8fold_TaCo2Te2} degenerate unconventional fermions. These new fermions are protected by symmetry elements and can occur at high symmetry points in the Brillouin region. Their degeneracy is related to irreducible representations (irreps) of the little group at high-symmetry points in the Brillouin Zone (BZ) for the space groups (SGs)~\cite{HSM_Sym_Bernevig}. These newly discovered fermions in solid-state systems not only contribute significantly to the advancement of fundamental science but also hold promising potential for the development of new materials.

\begin{figure}[t]
	\includegraphics[width=0.47\textwidth]{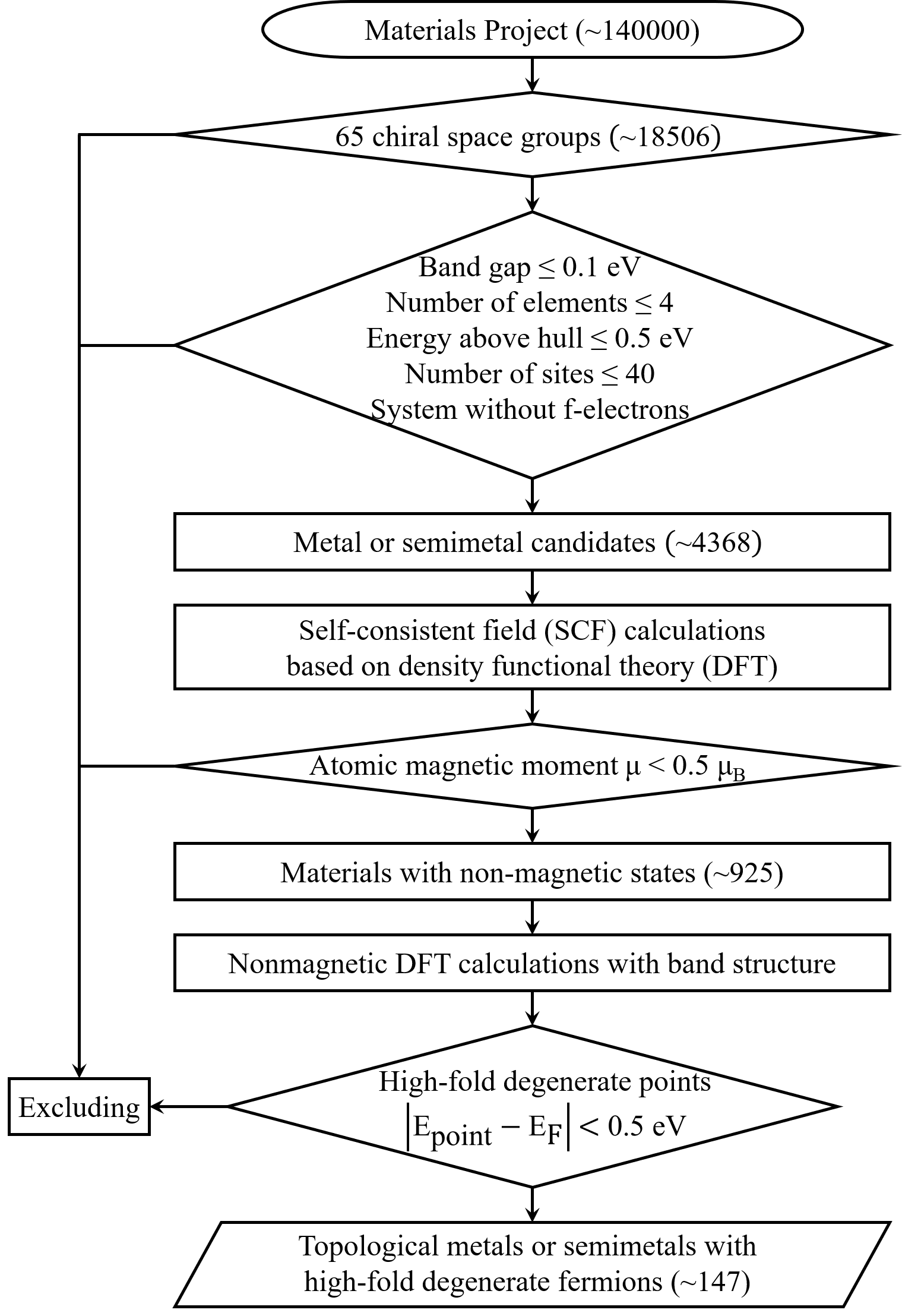}
	\caption{\textbf{High-throughput calculation process.}
		First, suitable candidates were screened through the database. Then, we performed SCF calculation to exclude the magnetic system. Finally, band calculation was carried out to find the high-fold degenerate points.}
	\label{fig:workflow}
	\vspace{-0.4cm}
\end{figure}

Here we are especially interested in chiral high-fold degenerate semimetals, which combine the advantages of both topological semimetals and chiral structures~\cite{HSM_chiral_AlPt_2019, TSM_chiral_Hasan_2021}. If an object is chiral, then it cannot have a symmetry plane, an inversion center, or a rotation-reflection axis ~\cite{Chiral_SolidState_2022}. Chiral compounds, which have various symmetry-dependent properties, are of particular interest. Related research has been carried out in mathematics~\cite{Chiral_Mathematics_1997}, physics~\cite{Chiral_NegativeRefraction_2004, Chiral_Skyrmion_2009}, biology~\cite{Chiral_Evolution_2005} and other fields. A series of chiral high-fold degenerate semi-metals have been discovered, such as CoSi family of materials~\cite{HSM_CoSi_2017,HSM_CoSi_2018,HSM_CoSi_chiral_2019,HSM_CoSi_2019},which is a near-ideal topological semimetal and includes representative materials like CoSi, AlPt and others.

\begin{figure*}[t]
	\centering
	\includegraphics[width=0.9\textwidth]{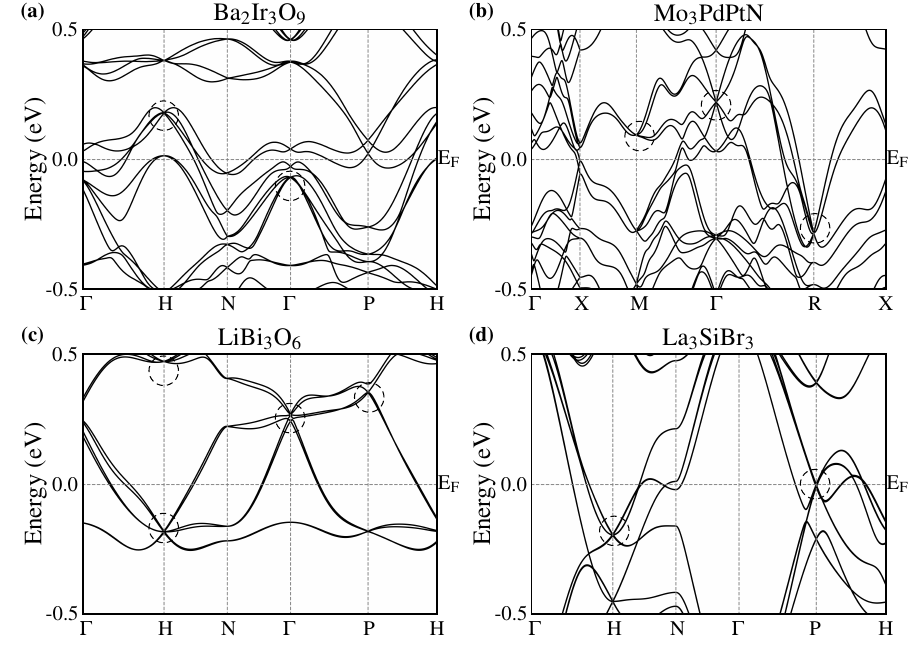}
	\vspace{-0.2cm}
	\caption{The band structure of (a) Ba$_2$Ir$_3$O$_9$ (SG 197), (b) Mo$_3$PdPtN (SG  198), (c) LiBi$_3$O$_6$ (SG 199) and (d) La$_3$SiBr$_3$ (SG 214). The horizontal gray dashed line indicates the Fermi energy,which is set to zero. The black dashed circles indicate the high-fold degenerate points.}
	\label{fig:band}
	\vspace{-0.4cm}
\end{figure*}

Many representative advantages of chiral high-fold degenerate semimetals can be found in the CoSi family of materials: high-fold chiral fermions with large Chern number, giant Fermi arcs across the entire surface Brillouin region and larger topological energy window that benefits from the absence of mirror symmetry. Moreover, special properties such as quantized circular photogalvanic effect~\cite{TSM_QCPE_2016,TSM_QCPE_2018}, gyrotropic magnetic effect~\cite{TSM_GME_2016}, and other special properties found in chiral high-fold degenerate semimetals make these promising in the fields of ferroelectricity, optical activity and so on. However, the most current research on chiral high-fold degenerate semi-metals is still focused on CoSi and its family of materials. It is necessary to discover more new materials for further theoretical and experimental research.

In this study, the high-throughput calculation and screening are carried out by classification of space groups from the perspective of crystal symmetry. 147 chiral materials with high-fold degenerate points have been found. Within these TSMs, not only three-, four-, and six-fold fermions have been found in high symmetry points, but also Weyl fermions are distributed in the Brillouin region. Multiple Fermi arcs that span the surface Brillouin zone can be observed. Our results confirm the existence of unconventional chiral fermions and conventional Weyl fermions in these chiral high-fold degenerate semi-metals, which makes it a promising platform for exploring the physical properties of chiral fermions.

\vspace{-0.3cm}
\section{Algorithm}
\vspace{-0.2cm}
We show the algorithm of high-throughput process in Figure~\ref{fig:workflow}. Starting with crystal structures from the material database Materials Project, we screened materials belonging to 65 chiral space group  that have structures that do not have an inversion center or mirror plane. To facilitate a fast diagnosis, we limited the number of sites to less than 40 and number of elements to less than 4. We are not interested in insulators, so materials in which the energy gap is greater than 0.1 eV were excluded. The energy above hull was limited to 0.5 eV to ensure the stability of structures. Lanthanides, except for lanthanum, and actinides were not considered due to their strong correlation. After filtering, we obtained 4368 candidate materials from the database. 

\begin{figure*}[t]
	\centering
	\includegraphics[width=0.9\textwidth]{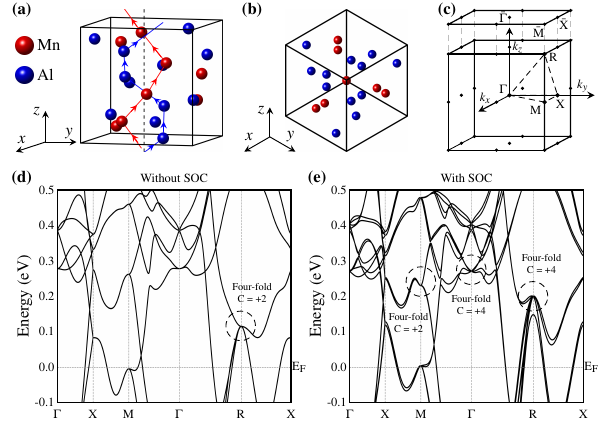}
	\vspace{-0.2cm}
	\caption{(Color online) \textbf{Crystal and electronic structure.}
		(a) The cubic lattice structure of Mn$_{2}$Al$_{3}$. The black dashed line represents a screw axis, while the red and blue solid lines show the screw relationship.
		(b) The crystal structure in (111) direction
		(c) The bulk BZ and the projected surface BZ for (001) surfaces.
		(d) and (e) Calculated band structure without and with SOC.}
	\label{fig:structure}
	\vspace{-0.4cm}
\end{figure*}

The self-consistent field (SCF) calculation based on DFT was carried out. Magnetic structures with atomic magnetic moment exceeding 0.5 $\mu_{B}$ were excluded according to the results. For the remaining 925 materials, we performed nonmagnetic DFT calculations with band structure to select high-fold points. We focused on the points that exist within 0.5 eV above and below Fermi energy because they greatly affect the electron transport properties of the materials. A total of 147 materials from 15 SGs that meet the screening limits have been found. We selected some materials with clean band structures at the Fermi level for the calculation of subsequent topological properties (Figure~\ref{fig:band}).

Here, we select Mn$_{2}$Al$_{3}$ with the SG 213 ($P4_{1}$32) as an example to analyze its topological properties . Mn$_{2}$Al$_{3}$ has a cubic crystal structure with 20 atoms in one unit cell, as shown in Figure~\ref{fig:structure}(a). It possesses 24 symmetry operations, which can be generated by four of them: one 3-fold rotation axis along (111) direction and three 2-fold screw axis along (001), (010) and (110) direction. The BZ is also a cubic structure with four high-symmetry momentum points: the cube-centered $\Gamma$, the face-centered X, the edge-centered M and the corner R. The 2D BZ of the projected (001) surface and 3D BZ are shown in Figure~\ref{fig:structure}(c). 

\vspace{-0.8cm}
\section{Method}
\vspace{-0.2cm}
To investigate the electronic structures and properties, first-principles calculations in this work were carried out based on density functional theory (DFT) using the package Vienna Ab initio Simulation Package (VASP)~\cite{VASP}. 
Exchange-correlation potential was treated within the generalized gradient approximation (GGA) of the Perdew-Burke-Ernzerhof functional (PBE)~\cite{PBE}. The Monkhorst-Pack $k$ mesh of the BZ in the self-consistent process was taken as the grid of 6 $\times$ 6 $\times$ 6. For the calculation of selected materials, the $k$-point grid was 8 $\times$ 8 $\times$ 8. All calculations were made considering spin-orbit coupling (SOC). To study topological properties, the maximally localized Wannier functions (MLWF)~\cite{wannier90} and a tight-binding Hamiltonian based on the MLWF overlap matrix were generated by wannier90. Using the WannierTools package, we calculated SSs with the Green’s function method~\cite{Green_functon_1984, Green_functon_1985}
in a half-infinite boundary condition. Other properties like topological invariants and Berry curvature were also be calculated by the WannierTools package~\cite{WannierTools}.

\begin{figure*}[t]
	\centering
	\includegraphics[width=1\textwidth]{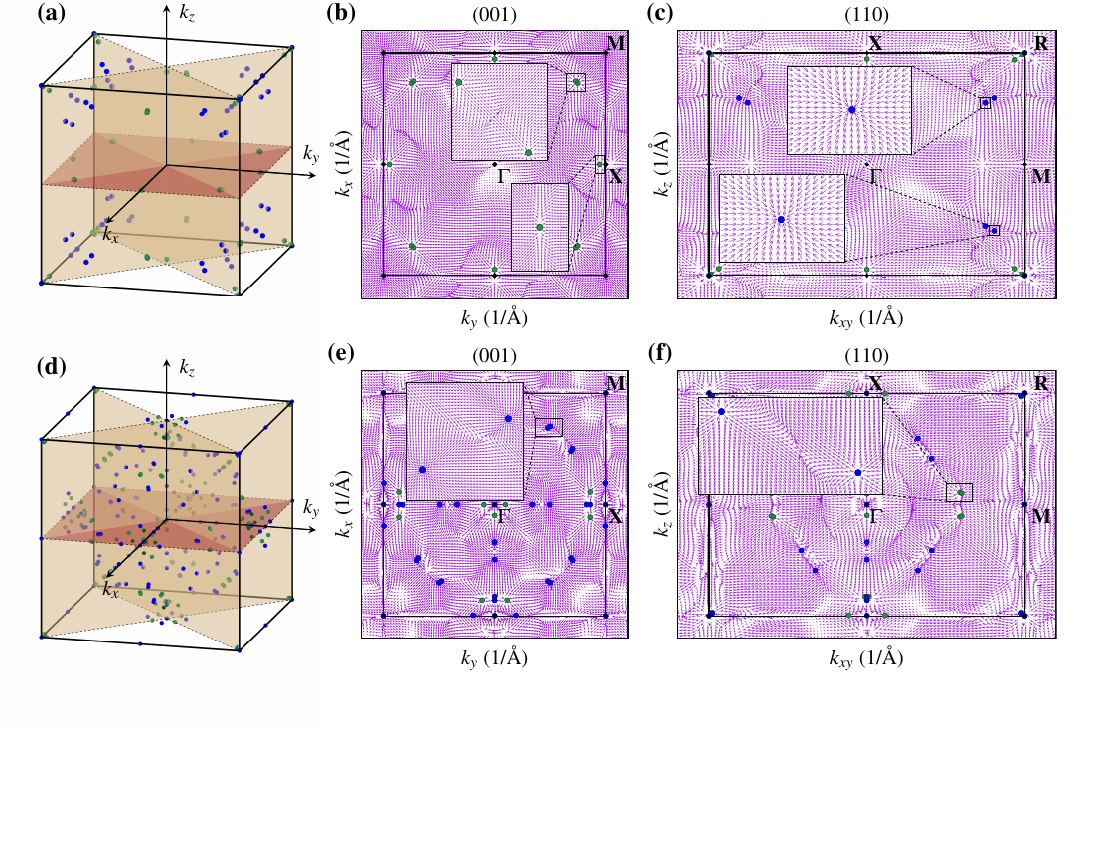}
	\vspace{-3cm}
	\caption{(Color online) 
		(a) The locations of the Weyl points (blue points represent the positive chirality and the green one represent negative) with the occupied number of 92. The crystal faces of (110), (1-10) and (001) are delineated by brown and cyan shaded planes.
		(b) and (c) Berry curvature of the $z=0$ and $x=y$ plane with the occupied number of 92. The high symmetry points are marked with black diamonds.
		(d) The locations of the Weyl points with the occupied number of 94.
		(e) and (f) Berry curvature of the $z=0$ and $x=y$ plane with the occupied number of 94.
	}
	\label{fig:berry}
	\vspace{-0.3cm}
\end{figure*}

\vspace{-0.6cm}
\section{Result}
\vspace{-0.2cm}

The dimension of irreducible representations of the little group of lattice symmetries at high-symmetry points in BZ corresponds to the degeneracy. So, we can classify materials based on their SGs and the little groups of the points to use symmetry to judge high-fold degenerate points by symmetry. Of the 147 materials found in previous calculation, we found high-fold degenerate points with degeneracies of three, four, and six. No eight-fold degenerate points were found, probably because of the lack of mirror symmetry. 

Figure~\ref{fig:structure}(d) and (e) show the band structure of Mn$_{2}$Al$_{3}$ without and with SOC, respectively. Without considering the SOC, a gapless point with four-fold degeneracy is observed about 0.12 eV above the Fermi level at the R point. This point is a monopole possessing topological charge +2. We also noticed the presence of a three-fold point above the Fermi level roughly 0.39 eV at the center of the BZ. 

After adding SOC, Figure~\ref{fig:structure}(e) clearly shows the change in the bulk band structure of Mn$_{2}$Al$_{3}$. For an arbitrary non-time reversal (TR) invariant point, the addition of the SOC term can lift its degeneracy due to the lack of inversion symmetry. At the $\Gamma$ point, the degeneracy of the original crossing doubles to form a four-fold points while the above six-fold point formed by the three-fold point splits into a two-fold crossing and a four-fold crossing whose topological charge is +4. Meanwhile, a six-fold degenerate point is found at the R point with the topological charge of +4, originating from the previous four-fold point. The X point also exhibites a four-fold crossing with the topological charge of +2. For the R and X point, which are on the boundaries of the 3D BZ, their degeneracy is protected by TR and nonsymmorphic symmetries. These fermions can have the largest possible separation allowed in crystals, which is characterized by previous studies on unconventional fermions. According to the Nielsen-Ninomiya theorem, the total chirality in 3D BZ must be zero. So, there must be other fermions with negative topological charge to cancel out these high-fold degenerate points.

\begin{figure*}[t]
	\centering
	\includegraphics[width=\textwidth]{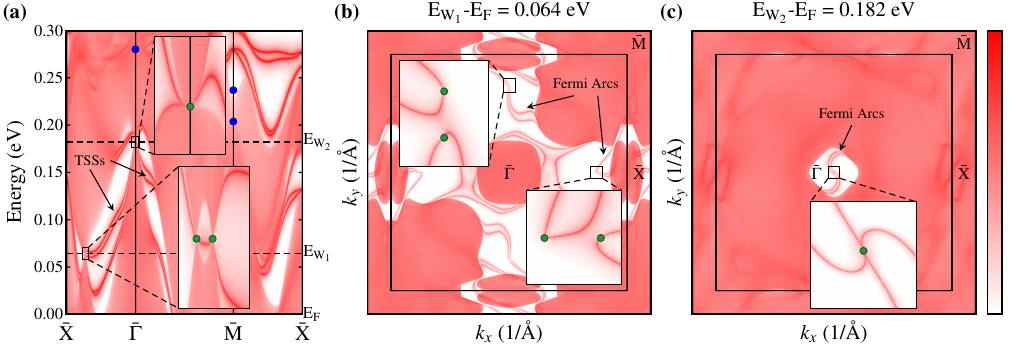}
	\vspace{-0.4cm}
	\caption{(Color online) \textbf{The (001) SSs of Mn$_2$Al$_3$.}
		(a) Energy dispersion for (001) surface along high symmetry lines. The calculated Fermi level is set to zero. High-fold points and Weyl points are marked with solid dots (blue points represent the positive chirality and the green ones represent negative).
		(b) and (c) The Fermi surface contours on the (001) surface at different energies.}
	\label{fig:ss}
	\vspace{-0.4cm}
\end{figure*}

Considering the distribution of the previous points, we focus on the bands with occupied number of 92 (about 0.1 eV above the Fermi energy) and 94 (about 0.2 eV above the Fermi energy). If one point can be obtained from another point by a symmetric transformation of the SG, then we call these two points a class. In addition to the fermions mentioned above, 6 and 18 classes of Weyl fermions have been respectively found, whose distribution in the BZ is shown in Figure~\ref{fig:berry}(a) and (d). Since the absence of mirror plane and inversion symmetry, these Weyl fermions do not appear in pairs of opposite topological charge. We analyze the distribution with the occupied number of 92 as an example. One class of Weyl fermions exists along ${\Gamma}-$X lines with the same topological charge of -2, occurring 6 times in 3D BZ. Two other classes are observed along ${\Gamma}-$R lines, with the topological charge of -1. They are invariant under three-fold rotation and counted 8 times. In addition, three classes of fermions do not exist along any symmetry-invariant axes. One of them has the charge of -1 and the other two are +1, all counted 24 times. We can find that these Weyl points are all distributed on the (001) and (110) crystal faces (and their symmetric equivalent faces). This distribution is related to the screw symmetry in SG 213. With the six-fold points at R, the total charge of BZ is zero. The same pattern can be found for Weyl points generated by the band with the occupied number of 94. 

If a sphere encloses a Weyl fermion, by Stokes theorem, the total integral of the Berry curvature over this closed torus must equal the total “monopole charge” carried by the Weyl fermion enclosed inside. For Mn$_{2}$Al$_{3}$, we calculated the Berry curvature of the (001) and (1$\bar{1}$0) crystal faces to observe the location of each Weyl point. Figure~\ref{fig:berry}(b)-(c) and~\ref{fig:berry}(e)-(f) show the Berry curvature of the above two crystal faces with the occupied number of 92 and 94, respectively. One can see that fermions with positive and negative chirality correspond to the “drain” and “source” points of “the magnetic field” in momentum space. 

TSSs with Fermi arcs connecting the projected Weyl fermions are characteristic of Weyl semimetals. For Mn$_{2}$Al$_{3}$, we calculate its SSs from (001) directions to demonstrate the exotic physics of topological fermions. The calculated results are shown in Figure~\ref{fig:ss} together with the FS plots. Along the ${\Gamma}-$X and ${\Gamma}-$M lines, topologically nontrivial SSs can be clearly observed, which emerge from the projected Weyl fermions and end at projections of bulk states. Unfortunately, the SSs connected to the high-fold degenerate points are so strongly coupled with the bulk states that it is hard to observe and distinguish. Instead, we show Fermi surface contours at two energies where the Weyl fermions can be observed. At E$_1$ [Fig~\ref{fig:ss}(b)], all the projected points on the (001) surface are generated either by two Weyl points with the same chirality -1, and their total chirality is -2. Correspondingly, two Fermi arcs are observed to extend from each Weyl point. Similarly, at E$_2$, two Weyl points with the chirality of -2 are projected into the ${\Gamma}$ point. Four Fermi arcs emerge from the G point and disappear into the bulk states.

\section{Summary}
\vspace{-0.1cm}
Although research on various TSMs is becoming more and more complete, most of the materials discovered so far are concentrated in the achiral structures. Many chiral topological semimetals discovered at present have high-fold points. To summarize and predict the possible existence of chiral high-fold degenerate semimetals, in this work, we performed high-throughput screening over all chiral structures from material databases. As a result of screening, 147 materials with high-fold degenerate points around the Fermi level have been found. We take Mn$_{2}$Al$_{3}$ as an example to show its longer Fermi arcs and larger topological energy window. Experiments have shown that these characteristics can bring greater carrier mobility and density, which are positive to some chemistry catalytic reaction. Thus, it is a promising field where chiral high-fold degenerate semi-metals be researched as surface catalysts.

\normalem
\bibliographystyle{aapmrev4-2}
\bibliography{article.bib}

\end{document}